\documentclass[aps,preprint,12pt,tightenlines]{revtex4}%
\usepackage{amsfonts}
\usepackage{amsmath}
\usepackage{amssymb}
\usepackage{graphicx}
\usepackage{color}%
\providecommand{\U}[1]{\protect\rule{.1in}{.1in}}
\begin{document}
\preprint{ }
\title[Quantum Noise and Quantum Vacuum]{Quantum Noise in Balanced Differential Measurements in Optics: Implication to
the Wave Modes of Quantum Vacuum}
\author{C. S. Unnikrishnan}
\affiliation{Tata Institute of Fundamental Research, Homi Bhabha Road, Mumbai - 400005, India}
\email{unni@tifr.res.in}

\begin{abstract}
\medskip Experimental tests for assessing the physical reality of the
hypothetical wave modes of quantum vacuum with zero-point energy are of
fundamental importance for quantum field theories and cosmology. Physical
effects like the Casimir effect have alternate description in terms of
retarded interaction between charged matter, due to quantum fluctuations of
material dipoles. However, there are simple quantum optical configurations
where the hypothetical quantum vacuum modes seem to assume an essential real
role in the observable quantum noise of optical signals. I present the logical
and theoretical basis of a decisive test that relies on the comparisons of
balanced homodyne detection with a novel differential scheme of balanced
wave-front division detection, when the two real optical beams at the
detectors are derived from one coherent beam as input. Both ideal and
practical configurations of my experimental test are discussed. \ Results from
the experiments on balanced detection, beam localization of optical beams, and
atomic Bose-Einstein condensates are used to reach definite conclusions
against the reality of the wave modes of quantum vacuum. It is shown that the
entire quantum noise follows consistently from the state reduction of quantum
superpositions of particle-number states at the point of detection, where the
quantum measurement is completed. This is consistent with the demonstrated
applications of squeezed light in interferometry and quantum metrology. This
result achieves consistency between quantum noise in quantum optics and
observational cosmology based on general relativity, by avoiding the wave
modes of quantum vacuum with divergent zero-point energy density.
Generalization from the limited sphere of quantum optics to general quantum
field theories promises the complete solution to the problem of a divergent
cosmological constant.

\end{abstract}
\startpage{1}
\endpage{102}
\maketitle
\tableofcontents

\section{Introduction}

Quantum Optics as a mature theory is much younger than quantum mechanics and
quantum field theories. The new foundations were firmly established in 1963,
with papers from R. Glauber \cite{Glauber} and E. C. G. Sudarshan
\cite{ECG-1963}. There are different approaches to the calculations, which
have different foundational basis. One important issue is the notion of
`Quantum Vacuum'. In the quantum optical picture that quantizes
electromagnetic wave modes, the vacuum mode is a wave mode at frequency $\nu$
with the zero-point energy $h\nu/2$. In this picture, photons are just
excitations of the wave mode with different amplitudes, with the wave-energy
quantized in units of $h\nu$. That is, the wave modes with finite zero-point
energy (ZPE) are present even if there is no detectable light. The use of real
wave mode subject to quantization rules in calculations raises the question of
the physical reality and observability of the hypothetical quantum vacuum
modes with a nonzero ZPE. The notion is seriously troublesome because each
mode has the ZPE of $h\nu/2$ and there is an infinity of such modes in space
and time, implying a divergent energy density that is in conflict with general
relativistic (observational) cosmology. The problem of the divergent ZPE
density, commonly called the problem of a divergent cosmological constant, has
remained unresolved for a long time \cite{Weinberg89,Rugh2002}.

In contrast, there is a logically robust picture that rejects real waves
underlying the quanta, pioneered and argued for by S. N. Bose, of photons as
particles of light obeying Bosonic rules of collective behaviour
\cite{Bose1924}. Hailed by Einstein as a fundamentally new way of treating the
statistical behaviour of quanta, this has no divergence problems (as Bose
himself had exclaimed \cite{Unni-SnC}). The Fock number state approach
formalizes this line of thought and integrates it to the underlying notion of
a `field' \cite{Fock1932}. The quantum state can be then described by
labelling the number of photons in a particular state, as $\left\vert
n\right\rangle $ and forming arbitrary superpositions. The vacuum state is
just the formal no-photon state $\left\vert 0\right\rangle $. The theory of
quantum optics treats the vacuum state as fundamental because it serves as the
basis from which n-photon Fock states are `created' as $\left(  a^{\dag
}\right)  ^{n}\left\vert 0\right\rangle $. The superposition of such number
states then provides all states of light. Modern quantum optics uses a hybrid
view, based on quantum field theories, where the notion of the underlying
field and particles as excitations is retained. This is reinforced by the
usage of coherent states as a universal basis, pioneered by E. C. G. Sudarshan
\cite{ECG-1963}. However, while mathematically complete, this hybrid view
carries the same conceptual inconsistency of the divergences.

In the rigorous `Dirac view' of quantum mechanics, the fluctuations in
physical quantities happens at the (repeated) measurements and there is no
fluctuations in quantum state itself either in time or in space until a
measurement in made. In contrast, the description with the quantum vacuum
modes explicitly links the quantum noise with the space-time fluctuations of
the amplitude and phase of the modes. Thus, the coherent state in the
Dirac-Fock picture is the superposition of all number states at all instants,
until the measurement at a detector. In the `mode-view', in contrast, the
field fluctuates in its amplitude and phase, the statistical ensemble average
of which gives the familiar `stick-and-ball' representation of the quantum
coherent state and its uncertainty noise. The latter view is obviously not
fully consistent with the principles of quantum mechanics, since it
\emph{implies definite amplitude and phase at each instant}, fluctuating. But
it is extensively in use due to its convenience.

There is another conceptual issue in the description of the quantum vacuum as
a physical state with average ZPE $h\nu/2$, carrying fluctuations in amplitude
and phase distributed as a Gaussian about zero. The probability for a
fluctuation in amplitude that far exceeds the average ZPE and even $h\nu$ is
finite; this is in conflict with requirement that no photon shall be detected
in the vacuum state. Though I will not discuss this further, the difference
between the `wave modes' view and the `quanta-state' view becomes crucial in
the discussion of quantum noise, its interpretation, and the exact expressions
in specific physical situations.

The Casimir force, Lamb shift, and the spontaneous emission from excited atoms
are phenomena cited in support of the physical reality of quantum vacuum
modes. However, the Casimir force, the prime example, can be derived as the
interaction of the quantum fluctuating dipoles (atoms) in the material making
the two surfaces, as the retarded van der Waals force, or as the integrated
Casimir-Polder force between a material boundary and an atom
\cite{Schwinger,Milton,Milonni}. Since all real boundaries are equal to the
factual presence of matter with quantum zero-point motion, and not mere static
mathematical conditions, the necessity of the vacuum modes in the Casimir
force cannot be insisted because that would be double-counting; either picture
is mathematically consistent when invoked alone. In other words, when wave
modes and their differential radiation pressure are invoked to calculate the
Casimir effect, one has to assume passive mathematical boundaries without any
atomic and electromagnetic structure subject to quantum mechanical zero-point
fluctuations. However, \emph{since matter and its zero point fluctuations are
the only reality that is directly observed and verified, the mode picture can
be seen only as a calculational tool, without physical reality to the wave
modes}. This argument is logically robust and it solves the problem of the
divergent ZPE, since the matter density in the universe is finite and its ZPE
is negligible. Therefore, there is strong reason and motivation to explore
laboratory physical situations to find a direct demarcating experiment that
tests the physical reality or otherwise of quantum vacuum modes. The rest of
the paper is devoted to the identification of such experiments, results, and
their interpretation.%

%TCIMACRO{\FRAME{ftbpFU}{2.8543in}{2.45in}{0pt}{\Qcb{The scheme for the
%balanced homodyne detection. The differential signal after the differencing
%circuit and its fluctuations are directly measured.}}{\Qlb{f-one}}%
%{fig1.eps}{\special{ language "Scientific Word";  type "GRAPHIC";
%width 2.8543in;  height 2.45in;  depth 0pt;  original-width 3.1208in;
%original-height 2.6758in;  cropleft "0";  croptop "1";  cropright "1";
%cropbottom "0";  filename 'fig1.eps';file-properties "XNPEU";}}}%
%BeginExpansion
\begin{figure}
[ptb]
\begin{center}
\includegraphics[
height=2.45in,
width=2.8543in
]%
{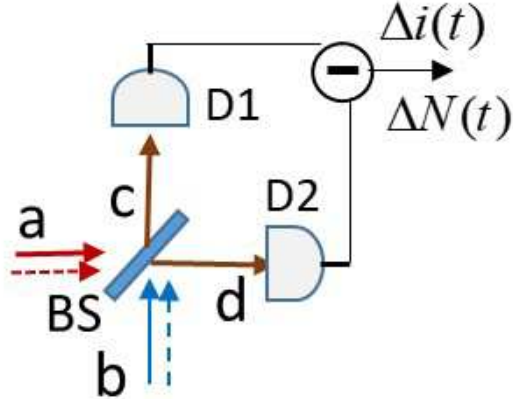}%
\caption{The scheme for the balanced homodyne detection. The differential
signal after the differencing circuit and its fluctuations are directly
measured.}%
\label{f-one}%
\end{center}
\end{figure}
%EndExpansion

The simple physical situation involving a `beam splitter' (BS) in
interferometers and quantum optical experiments is often discussed as a
reliable indication of the necessity of the real quantum vacuum mode. The
two-port balanced homodyne detection (BHD) of light \cite{Yuen-Chan} exploits
the fundamental peculiarity of the beam splitter that there is relative phase
of $\exp(i\pi)=-1$ between the two output beams of the passive device (figure
\ref{f-one}). BHD involves a strong phase coherent local oscillator (LO) beam
of amplitude $\alpha+\delta\alpha$ at one port (say 'a' port) of a 50:50 beam
splitter and the weak signal beam $s+\delta s$ at the other. The terms
$\delta\alpha$ and $\delta s$ represent the fluctuations in the fields. The
beams emerge superposed from the two output ports and the intensity is
measured in each port independently by photo-detectors. The difference in the
outputs of the detectors is obtained after a differencing circuit, and this is
the homodyne signal, which consists of only the interference terms $\left(
\alpha+\delta\alpha\right)  \left(  s+\delta s\right)  $ between the fields in
the two input ports due to the crucial relative phase factor of $\pi$ at the
BS \cite{Schumaker83,Loudon-paper,Loudon,Chiao}. If we consider the situation
when only the LO light is entering one port, with no light at the other, one
would have expected a zero homodyne signal because the cross interference term
would be identically zero. Yet, the actual homodyne output is non-zero
Gaussian noise centered on mean zero. This is the reason to postulate that a
physically real fluctuating quantum vacuum mode ($s=0,\delta s\neq0$) enters
the open empty port of the beam splitter and mixes (interferes) with the LO to
give the observed noise in the homodyne output. This interpretation was first
stressed by C. M. Caves in the context of the quantum noise in the
interferometric gravitational wave (GW) detectors based on Michelson
interferometers with the balanced beam splitter
\cite{CavesPRL80,CavesPR81,Schumaker-Caves85}. The radiation pressure noise is
proportional to difference in the mode intensities in the two arms, $\left(
c^{\dag}c-d^{\dag}d\right)  $, which can be written in terms of the
interference terms of the input operators $a_{1}$ and $a_{2}$ as $\left(
a^{\dag}b+b^{\dag}a\right)  $. This is also the \ difference signal measured
in the homodyne scheme. Then, the vacuum mode through the open port of the
beam splitter enters the description even if there is no input light in it. At
present, this is the standard interpretation for the irreducible quantum noise
in the balanced homodyne output. More over, the demonstrated use of squeezed
light in GW detectors is based on this physical picture
\cite{LIGO-Squeeze,GW-Squeeze}. Thus, quantum optics involving a beam splitter
seems to provide a ubiquitous situation and transparent proof for the physical
reality of the quantum vacuum mode. However, the divergent ZPE in general
relativistic cosmology remains as a genuine and troublesome physical problem.

How does one reconcile the two contradictory physical pictures, each with
seemingly convincing support? To obtain a reliable answer experimentally, I
devised a novel \emph{signal differencing configuration that avoids the phase
factor }$\pi$ of the beam splitter, which can then be compared with the
homodyne scheme. My results are completely consistent with Fock-state quantum
mechanics without the real wave modes of quantum vacuum. Then, quantum noise
is explained as entirely due to the reduction of the quantum state where the
quantum measurement is completed right at the detector, with \emph{no physical
role for the differencing operation after the intensity detectors}. While
resolving the conflict with general relativistic cosmology, my results are in
consistency with demonstrated quantum physical applications like squeezed
light metrology.

\section{Balanced Differential Measurements}

The simple physical situation involving a 50:50 `symmetric' beam splitter in
interferometers and quantum optical experiments is conventionally interpreted
as a reliable indication of the necessity of the real quantum vacuum mode. See
figure 2A. One mode of real radiation $a$ is split into two $\left(
c,d\right)  $ in amplitude by the BS and the mode intensities in the two beams
after BS are $c^{\dag}c$ and $d^{\dag}d$. There is no real optical bean in
port `b'. When detected with differencing photo-detector, the quantity
measured directly is $\Delta N(t)=N_{c}-N_{d}=c^{\dag}c-d^{\dag}d,$ which can
be formally written in terms of the interference terms of the input mode
operators $a$ and $b$ as $\left(  a^{\dag}b+b^{\dag}a\right)  $. The explicit
calculations are described later. This relation between the input and output
beams can be traced to the fundamental feature of the BS that there is a
relative phase of $\pi$ between the two output beams of the BS. Due to this
factor, only the cross interference terms survive without cancellation in the
balanced subtraction $\Delta N(t)=\left(  c^{\dag}c-d^{\dag}d\right)  $. This
is the signal measured in the balanced homodyne scheme. But there is no real
beam in the port `b'! However, when such an experiment is done, one does not
measure `zero' at the detector output. The average differential signal
$\left\langle \Delta N(t)\right\rangle =0$ and there is irreducible quantum
noise and variance in $\Delta N(t)$. This is the quantum noise, consistent
with the (non)commutation relations $\left[  a,a^{\dag}\right]  =\left[
b,b^{\dag}\right]  =1$. Thus, \emph{one is forced to postulate that the
invisible quantum vacuum mode enters through the open port `b' of the beam
splitter}, even if there is no input light in it. Only then the quantity
$\left(  a^{\dag}b+b^{\dag}a\right)  $ can be non-zero. At present, this is
the standard interpretation for the irreducible quantum noise in the balanced
homodyne output.%

%TCIMACRO{\FRAME{ftbpFU}{6.4134in}{1.467in}{0pt}{\Qcb{The two differential
%detection schemes to test the physical reality of the quantum vacuum modes.
%The dashed arrows indicate the hypothetical vacuum modes. Left: Balanced
%homodyne detection (BHD) with one real optical input at port `a'. The
%differential photodetector gives either the differential current $\Delta i(t)$
%or the differential photon counts $\Delta N(t)$. Right: Balanced
%wavefront-division detection (BWDD). The beam is expanded and then the
%wavefront is split equally into two parts and sent to the differential
%detector. }}{\Qlb{f-two}}{fig2.eps}{\special{ language "Scientific Word";
%type "GRAPHIC";  maintain-aspect-ratio TRUE;  display "USEDEF";
%valid_file "F";  width 6.4134in;  height 1.467in;  depth 0pt;
%original-width 5.5749in;  original-height 1.2536in;  cropleft "0";
%filename 'fig2.eps';file-properties "XNPEU";}}}%
%BeginExpansion
\begin{figure}
[ptb]
\begin{center}
\includegraphics[
height=1.467in,
width=6.4134in
]%
{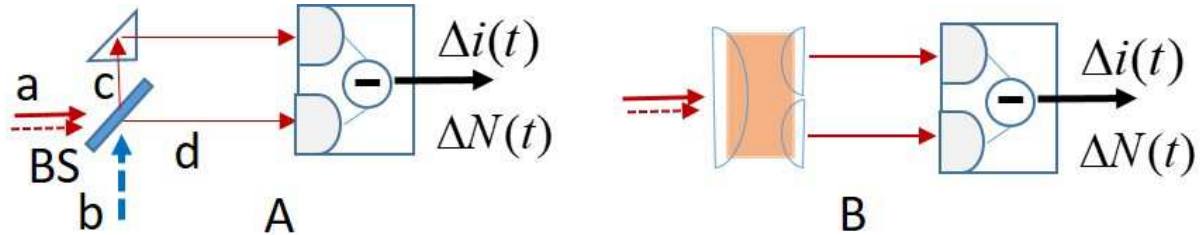}%
\caption{The two differential detection schemes to test the physical reality
of the quantum vacuum modes. The dashed arrows indicate the hypothetical
vacuum modes. Left: Balanced homodyne detection (BHD) with one real optical
input at port `a'. The differential photodetector gives either the
differential current $\Delta i(t)$ or the differential photon counts $\Delta
N(t)$. Right: Balanced wavefront-division detection (BWDD). The beam is
expanded and then the wavefront is split equally into two parts and sent to
the differential detector. }%
\label{f-two}%
\end{center}
\end{figure}
%EndExpansion

How reliable is this quantum optical evidence? Can one find an equally simple
measurement scheme in quantum optics that invalidates this interpretation and
disproves the physical reality of the invisible quantum vacuum mode? This is
indeed possible with a \emph{simple differential detection scheme where one
splits the wavefront, instead of the amplitude,} of the coherent beam before
the detection with the differential detector. I call this the balanced
wavefront-division detection (BWDD).  The detected quantity in BWDD is also
the difference in the intensity in the two detectors $\left(  \Delta
N(t)=a^{\dag}a-a^{\prime\dag}a\right)  $ where $a^{\prime}$ is the symmetric
replica of the mode $a$, being split from the same wavefront with the same
mode features (figure 2B). Both parts of the wavefront have quantum noise,
carried by the same quantum vacuum mode associated with the $a$ beam. However,
\emph{since the crucial phase factor }$\pi$\emph{ is absent in the
wavefront-split differential scheme, the subtraction is total, with no
surviving interference term}. Thus, the same quantum optical calculations that
attributed the quantum noise entirely to the uncanceled interference terms in
BHD predict near-zero noise at the differential output of BWDD. This
difference between the two schemes is the vital idea of the experiment. Yet,
from the quantum mechanical analysis of a coherent beam as an instantaneous
superposition of all particle (photon) number states, instead of as a field
mode with noise fluctuations, the difference signal and its noise variance
should not be zero; rather, the noise in $\Delta N$ should be $\sqrt{\bar
{N}_{c}+\bar{N}_{d}}$ in both BHD and BWDD, because of the completion of the
quantum measurement right at the detectors. It is this difference in the
predicted residual noise in the differential signal that helps reject the
reality of the hypothetical quantum vacuum modes.

\section{Differential Signals and Their Noise}

\subsection{Direct Detection of a Coherent Beam}

First, let us consider the quantum noise in a single mode of coherent
radiation, detected with a photodetector. The detected intensity is related to
the normal ordered number operator $a^{\dag}a$. In terms of the coherent
amplitude $\alpha$ this is just $\alpha\alpha^{\ast}$. The variance needs the
square of the number operator, $N^{2}=a^{\dag}aa^{\dag}a$, which is not a
normal ordered operator product. However, $a^{\dag}aa^{\dag}a=$ $a^{\dag
}\left(  a^{\dag}a+1\right)  a=\left\vert \alpha\right\vert ^{4}+\left\vert
\alpha\right\vert ^{2}$. \ Since $a^{\dag}a\gg1$ in the experiments we
consider, all quantum mechanical calculations are accurately done using the
complex coherent amplitude. The optical amplitude is $\alpha=a+\delta
a_{1}+i\delta a_{2}$, where the quantum fluctuations in the two quadratures
are designated as $\delta a_{1}$ and $\delta a_{2}$
\cite{Schumaker83,Loudon-paper}. \ For a coherent beam, the uncertainty
principle dictates $\left\vert \delta a_{1}\delta a_{2}\right\vert =1/4$. The
phase of the beam is fixed such that $a$ is real. We assume that there is no
other excess amplitude noise of the optical beam (like technical fluctuations
of the laser intensity). The detector output is
\begin{equation}
i(t)\propto N(t)=a^{\dag}a\simeq\alpha\alpha^{\ast}=a^{2}+2a\delta
a_{1}+\delta a_{1}^{2}+\delta a_{2}^{2}%
\end{equation}
The quantum noise is reflected in the last three terms, of which only the
first is significant in magnitude. Then,
\[
N(t)\simeq a^{2}+2a\delta a_{1}\rightarrow\bar{N}\pm\sqrt{\bar{N}}%
\]
The variance is
\begin{equation}
V(N)=\left\langle \left(  N-\bar{N}\right)  ^{2}\right\rangle \simeq
4a^{2}\delta a_{1}^{2}%
\end{equation}
We see that the entire quantum noise is consistently described by the
fluctuations $\delta a_{1}$ and $\delta a_{2}$ in the `quantum vacuum mode'.
\ When evaluated, $V(N)\simeq\bar{N}$, because $\delta a_{1}=\delta a_{2}=1/2$
for a coherent mode, from the uncertainty relation $\delta a_{1}\delta
a_{2}=1/4$.

The alternate and conceptually very different interpretation of quantum noise
based on a particle picture of standard quantum mechanics is also consistent.
There, the coherent state of photons is a superposition of all number states
$\left\vert n\right\rangle $ with Poissonian weights.
\[
\left\vert \alpha(t)\right\rangle =\exp(-\left\vert \alpha\right\vert ^{2}/2)%
%TCIMACRO{\dsum \limits_{n=0}^{n=\infty}}%
%BeginExpansion
{\displaystyle\sum\limits_{n=0}^{n=\infty}}
%EndExpansion
\frac{\alpha^{n}}{\sqrt{n!}}\left\vert n\right\rangle
\]
The detector detects the average $\bar{N}=\alpha^{2}$ and its fluctuations
(standard deviation) are $\sqrt{\bar{N}}$. \ Thus $V(N(t))=\bar{N}$. \ The
quantum measurement is completed at the square-law detector. In fact, it is
identical in structure to a coherent state of atoms (for which there is no
underlying real wave or vacuum mode with a zero-point energy). \emph{There is
no real wave mode and its fluctuations in this picture}. Then one can get
fluctuations in the photon number on detection, with Poissonian probability,
with the collapse of the state that happens (only) at the detection event.

For this particular case of the direct detection of the optical beam, either
view is consistent. Our goal is to demarcate and decide between the two
different views, by a decisive experiment.

\subsection{Balanced Homodyne Detection}

We refer to the figure \ref{f-two}A. The field amplitudes corresponding to the
modes are complex quantities. Since only relative phases are observable, one
of the modes can be taken as real quantity. The quantum noise in each mode is
however in two quadratures. So, $a\rightarrow a+\delta a_{1}+i\delta a_{2}$.
All experiments under discussion have only one real optical beam. For the BS
experiment, there are two input ports and two possible independent modes
(different `k' vectors etc.). But, since there is no real optical beam in the
second port, only the hypothetical and invisible quantum vacuum modes with
fluctuations enter the final expressions, $b\rightarrow\delta b_{1}+i\delta
b_{2}$.

The general beam splitter has the transmission $T=\left\vert t\right\vert
^{2}$, to port `d' from port `a', and hence reflection $R=\left\vert
r\right\vert ^{2}=1-T$ from port `a' to port `c' (or from port b to port d).
Fixing explicitly one relative phase of $\pi$ between ports `b' and `d', we
have real $r$ and $t$. The phase of `a' beam is chosen such that $a$ is real.
Since we are concerned with experiments on quantum vacuum, $b$ will eventually
be zero or very small. Thus, the terms second order in small quantities
($b^{2},\delta a^{2},\delta b^{2}\delta a\delta b,b\delta b$) are negligible.
The beam splitter combines the two input modes to give two output modes $c$
and $d$. The photon counts in port `c' (reflection of $b$ and transmission of
$a$) is
\begin{align}
N_{c} &  =c^{\dag}c=\left(  ra^{\dag}+tb^{\dag}\right)  \left(  ra+tb\right)
\simeq\left\vert r\left(  a+\delta a\right)  +t\left(  b+\delta b\right)
\right\vert ^{2}\label{N_c}\\
&  \left(  r\left(  a+\delta a\right)  +t\left(  b+\delta b\right)  \right)
\left(  r\left(  a+\delta a\right)  +t\left(  b+\delta b\right)  \right)
^{\ast}\nonumber\\
&  =R\left(  a^{2}+\delta a^{2}+2a\delta a\right)  +T\left(  b^{2}+\delta
b^{2}+2b\delta b\right)  \nonumber\\
&  +rt(ab^{\ast}+a\delta b^{\ast}+b\delta a+\delta a\delta b^{\ast
})+rt(ba+b\delta a^{\ast}+a\delta b+\delta b\delta a^{\ast})\nonumber\\
&  \simeq Ra^{2}+2Ra\delta a+\sqrt{TR}\left(  ab^{\ast}+ab+a\delta b^{\ast
}+a\delta b\right)  \label{N_cf}\\
&  =Ra^{2}+2Ra\delta a+2\sqrt{TR}\left(  ab_{1}+a\delta b_{1}\right)
\end{align}

For the other output port `d', inserting the relative phase factor $\pi$ from
port b to d,
\begin{align}
N_{d}  &  =d^{\dag}d\simeq\left[  t\left(  a+\delta a\right)  -r\left(
b+\delta b\right)  \right]  ^{2}\label{N_d}\\
&  \simeq Ta^{2}+2Ta\delta a-2\sqrt{TR}\left(  ab_{1}+a\delta b_{1}\right)
\end{align}

The symmetric (50:50) BS has equal transmission and reflection coefficients
($tt^{\ast}=T=1/2=R=rr^{\ast}$). Then,
\begin{align}
N_{c}  &  =\frac{1}{2}a^{2}+a\delta a+ab_{1}+a\delta b_{1}\\
N_{d}  &  =\frac{1}{2}a^{2}+a\delta a-ab_{1}-a\delta b_{2}%
\end{align}

Balanced homodyne detects the difference signal $\Delta N(t)=N_{c}-N_{d}$ and
its noise directly. The total quantum noise is $2\left(  a\delta b_{1}+\delta
a_{1}\delta b_{1}+\delta a_{2}\delta b_{2}\right)  $, where the last two terms
are negligible. \emph{The difference signal is identical to the (cross)
interference term between the two beams in the ports a and b},%

\begin{align}
\Delta N(t)  &  =2ab_{1}+2a\delta b_{1}\label{Diff_N}\\
\left\langle \Delta N\right\rangle  &  =2ab_{1}%
\end{align}
It is remarkable that we get the interference term in the difference signal
despite a `phase-destroying' square law detection at each detector. Similar
`magic' was seen in the classic case of the Hanburry Brown-Twiss intensity
interferometry that prompted the new theory of correlations and quantum
coherence in optics. Most importantly, the signal has no self interference
terms between the mode amplitude $a$ and its quantum vacuum fluctuations
$\delta a$. Only the cross interference term $a\delta b_{1}$ is present.
\emph{It should be noted that if the BS did not introduce the }$\pi$\emph{
phase, all signs in }$N_{d}=d^{\dag}d$ \emph{(eq. \ref{N_d}) would have been
positive and the difference signal in equation \ref{Diff_N} would have been
identically zero in the subtraction, without any averaging}.

When there is only one coherent beam at the port `a', with only `vacuum' at
the port `b', $\left\langle \Delta N\right\rangle =0$. \ The balanced homodyne
signal has the variance%
\begin{equation}
V(\Delta N(t))=\left\langle \left(  \Delta N-\overline{\Delta N}\right)
^{2}\right\rangle =4a^{2}\delta b_{1}^{2}+4\delta a_{1}^{2}\delta b_{1}%
^{2}+4\delta a_{2}^{2}\delta b_{2}^{2}+8\left\langle \delta a_{1}\delta
b_{1}\delta a_{2}\delta b_{2}\right\rangle \simeq4a^{2}\delta b_{1}%
^{2}\label{Var-homo}%
\end{equation}
All other terms average to zero. \ The last of the four terms can give a
nonzero average if there are correlations in the two quadratures of each mode.
However, only the first term is significant in magnitude. It is of order $N$,
whereas the other terms are of order $1$, totally negligible. Thus, $V(\Delta
N)\simeq4a^{2}\delta b_{1}^{2}$. Therefore, \emph{the quantum noise in
balanced homodyne detection is attributed entirely to the quantum noise in the
`b' mode, with the amplitude of the `a' mode beam acting as a noise-free
linear amplifier, with gain} $\bar{N}$.

It is instructive to look at the variance of the signal at each detector. We
have
\begin{align}
V(c)  &  =\left\langle \left(  N_{c}-\bar{N}_{c}\right)  ^{2}\right\rangle
=a^{2}\delta a_{1}^{2}+a^{2}\delta b_{1}^{2}+\delta a_{1}^{2}\delta b_{1}%
^{2}+\delta a_{2}^{2}\delta b_{2}^{2}+2\delta a_{1}\delta b_{1}\delta
a_{2}\delta b_{2}\simeq a^{2}\delta a_{1}^{2}+a^{2}\delta b_{1}^{2}\\
V(d)  &  =\left\langle \left(  N_{d}-\bar{N}_{d}\right)  ^{2}\right\rangle
=a^{2}\delta a_{1}^{2}+a^{2}\delta b_{1}^{2}+\delta a_{1}^{2}\delta b_{1}%
^{2}+\delta a_{2}^{2}\delta b_{2}^{2}+2\delta a_{1}\delta b_{1}\delta
a_{2}\delta b_{2}\simeq a^{2}\delta a_{1}^{2}+a^{2}\delta b_{1}^{2}%
\end{align}
\emph{Clearly the expression for the sum of the variances of the two signals,
}$V(c)+V(d)\simeq2\left(  a^{2}\delta a_{1}^{2}+a^{2}\delta b_{1}^{2}\right)
$\emph{ is not equal to the variance of the homodyne difference signal (eq.
\ref{Var-homo})}, $V(\Delta N(t)\simeq4a^{2}\delta b_{1}^{2}$. \ The
theoretical quantum noise in the output of each detector is contributed
equally by the quantum fluctuations $\delta a$ and $\delta b$ of the quantum
vacuum modes `$a$' and `$b$', whereas the theoretical quantum noise in the
direct difference signal is entirely contributed by the fluctuations in the
`$b$' mode, $\delta b$.
\begin{equation}
V(c)+V(d)\simeq2\left(  a^{2}\delta a_{1}^{2}+a^{2}\delta b_{1}^{2}\right)
\neq V(\Delta N(t)\simeq4a^{2}\delta b_{1}^{2}%
\end{equation}
\emph{This is an important and fundamental result}. \ For equal uncertainties
in both quadratures, the numerical magnitudes $2\left(  a^{2}\delta a_{1}%
^{2}+a^{2}\delta b_{1}^{2}\right)  $ and $4a^{2}\delta b_{1}^{2}$ are
obviously indistinguishably equal. However, the result becomes crucial, with
quantitatively different total noise, in experiments with squeezed light or in
those situations where there are quantum correlations. For example, with
$\delta b_{1}$ squeezed by a factor $\exp(-s)$,
\begin{equation}
V(c)+V(d)\simeq2\left(  a^{2}\delta a_{1}^{2}+e^{-2s}a^{2}\delta b_{1}%
^{2}\right)
\end{equation}
whereas the variance in the homodyne signal is very different,
\begin{equation}
V(\Delta N)\simeq4e^{-2s}a^{2}\delta b_{1}^{2}%
\end{equation}

Hence, $\left\vert V(\Delta N)\right\vert \neq\left\vert V(C)+V(C)\right\vert
$. We see right here some fundamental inconsistency and conflict in the notion
of real quantum vacuum modes with the idea of measurement in quantum
mechanics, but I will defer a detailed discussion on that and focus on the new
wavefront division experiment.

\subsection{Balanced Wavefront-Divison Detection}%

%TCIMACRO{\FRAME{ftbpFU}{5.8182in}{1.6903in}{0pt}{\Qcb{Two configurations of
%balanced wavefront-division detection. This does not involve a 50:50 BS and
%associated relative phase of $\pi$. B) Scheme with one or more beam steering
%elements, like a prism or a mirror.}}{\Qlb{f-three}}{fig3.eps}%
%width 5.8182in;  height 1.6903in;  depth 0pt;  original-width 5.3458in;
%original-height 1.5326in;  cropleft "0";  croptop "1";  cropright "1";
%cropbottom "0";  filename 'fig3.eps';file-properties "XNPEU";}}}%
%BeginExpansion
\begin{figure}
[ptb]
\begin{center}
\includegraphics[
height=1.6903in,
width=5.8182in
]%
{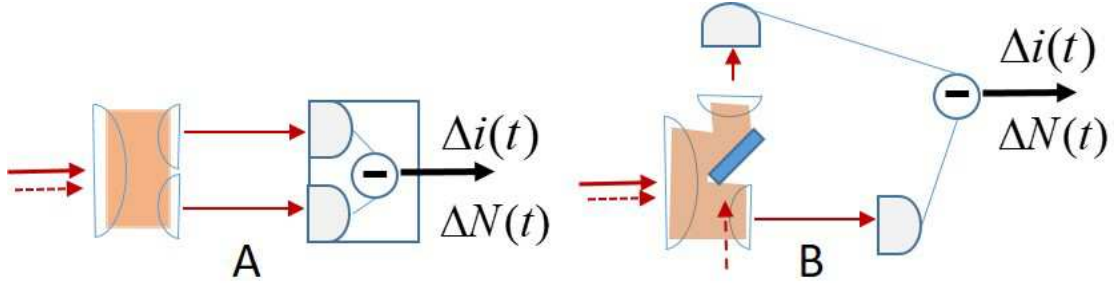}%
\caption{Two configurations of balanced wavefront-division detection. This
does not involve a 50:50 BS and associated relative phase of $\pi$. B) Scheme
with one or more beam steering elements, like a prism or a mirror.}%
\label{f-three}%
\end{center}
\end{figure}
%EndExpansion

We write the essential steps in the calculation for the balanced
wavefront-division differential detection (BWDD) schemes of figure
\ref{f-three}. The mode $a$ with its vacuum noise is split into two sections,
which we designate as $\alpha_{u}$ and $\alpha_{d}$. They are the same mode
with all quantum numbers matching. But we can try to distinguish the parts of
the wavefront with a spatial label, up (u) and down (d) of the wavefront.
\begin{align}
\alpha_{u} &  =\frac{1}{\sqrt{2}}\left(  a+\delta a_{1}+i\delta a_{2}\right)
\nonumber\\
\alpha_{d} &  =\frac{1}{\sqrt{2}}\left(  a+\delta a_{1}+i\delta a_{2}\right)
\end{align}
The fact that there is only one beam and one mode now, from which two beams
are generated by diving the wavefront, and there is no $\pi$ phase to
introduce a negative sign in any term, reflects in the differential output.
The detectors generate the intensity signals,
\begin{align}
\alpha_{u}^{\dag}\alpha_{u} &  \simeq\alpha_{u}\alpha_{u}^{\ast}=\frac{1}%
{2}\left(  a^{2}+2a\delta a_{1}+\delta a_{1}^{2}+\delta a_{2}^{2}\right)  \\
\alpha_{d}^{\dag}\alpha_{d} &  \simeq\alpha_{d}\alpha_{d}^{\ast}=\frac{1}%
{2}\left(  a^{2}+2a\delta a_{1}+\delta a_{1}^{2}+\delta a_{2}^{2}\right)
\end{align}
\emph{The output of the differential detector is identically zero, without any
averaging}.
\begin{equation}
\Delta N(t)=\alpha_{u}^{\dag}\alpha_{u}-\alpha_{d}^{\dag}\alpha_{d}=0
\end{equation}
We have no freedom in this calculation if we follow the same logic as what was
used in the homodyne calculation. There, the split occurs in the time domain
whereas here it occurs in the spatial domain; that is the only difference.
There are only self interference terms of the mode amplitude with its own
quantum fluctuations here, but  there is no cross interference term with the
fluctuations of another mode.

The variance of the output signal $\Delta N(t)$ is obviously zero, since
$\Delta N(t)$ itself is zero.
\begin{equation}
V(\Delta N(t))=\left\langle \left(  \Delta N-\overline{\Delta N}\right)
^{2}\right\rangle =0
\end{equation}
If the balance is not ideal, but in the ratio $P:Q$ with $P\simeq Q$, then the
expression modifies to
\begin{align}
\Delta N(t) &  =(P-Q)a^{\dag}a\ll a^{\dag}a\label{Del-N}\\
V(\Delta N(t)) &  =4(P-Q)^{2}a^{2}\delta a_{1}^{2}\ll\bar{N}\label{Var-N}%
\end{align}
Therefore, the experiment can decisively determine whether a real quantum
vacuum mode is the source of the quantum noise or whether it is the collapse
of the instantaneous quantum superposition of the particle number states. If
the difference signal and its variance is near zero, then it supports the view
that the quantum vacuum mode with the ZPE is the carrier of the quantum noise.
On the other hand, the collapse of the particle number superposition at the
detectors predicts that the variance is the sum of the variances at the
individual detectors, $V(\Delta N(t))=V(c)+V(d)=\bar{N}_{c}+\bar{N}_{d}$,
\emph{because the quantum measurement is completed (only) at the square law
photodetectors}. Therefore, noise variance that scales as the intensity
(average photon number) in BWDD would be reliable evidence against the real
wave modes of quantum vacuum.

One might think and hope that one could introduce new rules for the wave modes
to save the interpretation of the zero-point wave as the source of the quantum
noise. One possibility is to postulate that the fluctuations are cancelled in
the subtraction only when the amplitude is split, but not when the wavefront
is split. This would be the case if different spatial portions of the same
wavefront have different fluctuations, even after integrating over the
response time of the detectors. But, this arbitrary postulate does not solve
the problem. We repeat the calculation with the new postulate. The intensity
signals with different and independent quantum noise $\delta a$ and $\delta
a^{\prime}$ in the two haves of the wavefront are,
\begin{align}
\alpha_{1}^{\dag}\alpha_{1} &  \simeq\alpha_{1}\alpha_{1}^{\ast}=\frac{1}%
{2}\left(  a^{2}+2a\delta a_{1}+\delta a_{1}^{2}+\delta a_{2}^{2}\right)  \\
\alpha_{2}^{\dag}\alpha_{2} &  \simeq\alpha_{2}\alpha_{2}^{\ast}=\frac{1}%
{2}\left(  a^{2}+2a\delta a_{1}^{\prime}+\delta a_{1}^{\prime2}+\delta
a_{2}^{\prime2}\right)
\end{align}
After the subtraction,
\begin{align}
\Delta N(t) &  =\alpha_{1}\alpha_{1}^{\ast}-\alpha_{2}\alpha_{2}^{\ast}\simeq
a(\delta a_{1}-\delta a_{1}^{\prime})<2a\delta a_{1}\\
V(\Delta N) &  =a^{2}(\delta a_{1}-\delta a_{1}^{\prime})^{2}\simeq
a^{2}\left(  \delta a_{1}^{2}+\delta a_{1}^{\prime2}\right)  -2a^{2}%
\left\langle \delta a_{1}\delta a_{1}^{\prime}\right\rangle =2a^{2}\delta
a_{1}^{2}-2a^{2}\left\langle \delta a_{1}\delta a_{1}^{\prime}\right\rangle
<4a^{2}\delta a_{1}^{2}%
\end{align}
Therefore, $V_{BWD}=V_{BH}/2$ for independent quantum fluctuations in the two
parts of the wavefront. So, this patch repair does not match the variance that
we got for the homodyne scheme in magnitude, $V_{BWD}(\Delta N(t))<V_{BH}%
(\Delta N(t))\simeq4a^{2}\delta b_{1}^{2}$. When the fluctuations are fully
anticorrelated ($\delta a_{1}=-\delta a_{1}^{\prime}$), then $\left\langle
\delta a_{1}\delta a_{1}^{\prime}\right\rangle =-\delta a_{1}^{2}$, but one
cannot get this condition for every partition of the wavefront. If this were
the case, then the total intensity would be $\alpha_{1}\alpha_{1}^{\ast
}+\alpha_{2}\alpha_{2}^{\ast}=a^{2}+a\left(  \delta a_{1}+\delta a_{1}%
^{\prime}\right)  =a^{2}$, with no variance! \emph{The wavefront-division
detection can provide a strong proof against the wave modes of quantum vacuum}.

\subsection{Insignificance of Steering Mirrors}

It is convenient (and often necessary) to use one or more steering mirrors in
the experiments to get the beam to the detector location etc. While it is
intuitively obvious that a mirror with high reflectivity does not affect the
concept or the quantitative expressions of the experiment, it is perhaps
appropriate to prove this explicitly. (This clarifying calculation, obvious to
most, is included after seeing persistent confusion among some colleagues that
a single steering mirror interpreted as a beam splitter with very low
transmission is essential and crucial in the wavefront division experiment to
compare it with the homodyne experiment). A steering mirror of reflectivity
$R$ can be interpreted as a $R:T$ beam splitter, with $T\ll R\simeq1$. We note
from equation \ref{N_cf} that the use of the mirror alters the quantum noise
contribution in that beam slightly,
\begin{align}
N_{c}-\bar{N}_{c}  &  =2Ra\delta a+2\sqrt{RT}a\delta b\\
V(N_{c}(t))  &  \simeq4R^{2}a^{2}\delta a^{2}+4RTa^{2}\delta b^{2}
\label{Var-mirror}%
\end{align}

The differential signal and its variance in the wavefront-division experiment
with a steering mirror (fig. \ref{f-three}B), assuming 50:50 wavefront
division before the mirror, is calculated next. The already divided amplitude
$\left(  a+\delta a\right)  /\sqrt{2}$ becomes $r\left(  a+\delta a\right)
/\sqrt{2}$ after the $R:T$ mirror. This is superposed with $t\delta b$ when we
assume that the vacuum mode leaks in due to the small transmissivity $T$. I
include also the $\pi$ phase for reflection at the mirror. The intensity at
the detector is
\begin{equation}
N_{c}=\left(  \frac{e^{i\pi}r}{\sqrt{2}}\left(  a+\delta a\right)  +t\delta
b\right)  \left(  \frac{e^{i\pi}r}{\sqrt{2}}\left(  a+\delta a\right)
+t\delta b\right)  ^{\ast}\simeq\frac{R}{2}\left(  a^{2}+2a\delta
a_{1}\right)  -\frac{2rt}{\sqrt{2}}a\delta b_{1}%
\end{equation}
\emph{The phase factor at reflection drops out}. The other half of the
wavefront goes straight to the detector `d'. So,
\begin{align}
N_{d} &  \simeq\frac{1}{2}\left(  a^{2}+2a\delta a_{1}\right)  \nonumber\\
\Delta N(t) &  =N_{d}-N_{c}=\frac{1}{2}(1-R)\left(  a^{2}+2a\delta
a_{1}\right)  +\sqrt{2RT}a\delta b_{1}\nonumber\\
\left\langle \Delta N(t)\right\rangle  &  =\frac{1}{2}(1-R)a^{2}\\
V(\Delta N) &  =(1-R)^{2}a^{2}\delta a_{1}^{2}+2RTa^{2}\delta b^{2}%
\simeq2Ta^{2}\delta b^{2}=\left(  T/2\right)  V(\Delta N_{BHD}%
)\label{VarDN-mirror}%
\end{align}
As stated before, with $T<1\%$, the variance in the differential signal is
negligible compared to the variance in the balanced homodyne signal,
suppressed by the factor $T/2$. \emph{The phase }$\pi$\emph{ at the steering
mirror is irrelevant, unlike the crucial phase of }$\pi$\emph{ in the homodyne
BS. }Similar remarks apply if a prism (total internal reflection) is used for
steering the beam.

\section{The Real Source of the Quantum Noise}

Now I show that the real source of the quantum noise is the quantum state
reduction of the particle superposition of the real optical beam at the
detector that completes the quantum measurement, in all cases in quantum
optics, without exception. The coherent state in the example of direct beam
detection is
\begin{equation}
\left\vert \alpha(t)\right\rangle =\exp(-\left\vert \alpha\right\vert ^{2}/2)%
%TCIMACRO{\dsum \limits_{n=0}^{n=\infty}}%
%BeginExpansion
{\displaystyle\sum\limits_{n=0}^{n=\infty}}
%EndExpansion
\frac{\alpha^{n}}{\sqrt{n!}}\left\vert n\right\rangle
\end{equation}
Before reaching the detector, the quantum state is a superposition of all
number states, with its only time evolution reflected in an overall phase.
Till the state $\left\vert \alpha\right\rangle $ interacts with the square law
detector, subject to an interaction Hamiltonian, the state is stable and under
pure unitary evolution. \emph{The quantum measurement is completed at the
square-law detector, and only at that point}. This gives a value for the
number of photons with probability $p(n)$ dictated by the Poissonian weight of
the state with $n$ number of photons. It is clear that whether we make the
measurement on the beam by direct detection, or on any part of the beam by
splitting it in amplitude or wavefront, the state is the same, multiplied by
an overall numerical factor. The relative probabilities remain the same. For
an ensemble of identical quantum measurements with the detector, a sequence of
such results are obtained; then we get the average $\bar{N}=\alpha^{2}$ and
its statistical fluctuations, with standard deviation $\sqrt{\bar{N}}$. \ Thus
$V(N(t))=\bar{N}$. \ 

For balanced homodyne detection as well as for balanced wavefront division
detection, each detector completes the quantum measurement independently and
the rest of the operations are irrelevant for the quantum state. Each detector
measures the average $\bar{N}/2$ with the standard deviation $\sqrt{\bar{N}%
/2}$. Therefore, the difference signal is centred on zero, with standard
deviation $\sqrt{\bar{N}}$. \ \emph{This is the prediction for any balanced
dual detector measurement from the theory that rejects the physical wave mode
of quantum vacuum. Then the noise is attributed to the quantum measurement
that reduces a quantum superposition of number states to a particular number
state in each measurement event}. It does not make any difference in the
expressions for the quantum noise whether one performs the individual
measurements and then takes the difference, or whether \ the differential
signal is directly measured. \ This interpretation and prediction are
universally applicable in all cases, from direct detection to general
unbalanced detection of light. Since the prediction for the variance of the
balanced differential signal is very different in the quantum state view and
in the vacuum-waves view ($\bar{N}$ vs $0$), one can decisively determine the
correct physical picture directly from the experiment and rule on physical
reality of the wave modes of quantum vacuum.

\section{Experimental Results}

\subsection{Balanced Homodyne Measurements}

Balanced homodyne measurement (fig. \ref{f-two}A) is the primary tool for the
characterization of quantum noise in squeezed light. There are already several
experimental results that are consistent with the calculations described in
the section on BHD. Therefore, BHD measurements that measure the difference
signal and its variance are consistent with the hypothesis that a real wave
mode of quantum vacuum entering the open port of the BS is the source of the
entire quantum noise \cite{Raymer,Leonhardt}. \emph{The quantum noise
variances in the individual detectors are, however, contributed by the vacuum
modes entering both the input ports. The differencing operation after the
intensiy detectors cancels one contribution while doubling the other}. Other
caveats are briefly mentioned in section IIB.

\subsection{Balanced Wavefront-Division Measurements}

Now we examine the hypothesis of the quantum vacuum modes against the balanced
wavefront-division measurements. The experimental scheme is as indicated in
figure \ref{f-three}. Direct experiments can be with two matched
photodetectors, at relatively high intensity of about a mW (%
%TCIMACRO{\TEXTsymbol{>}}%
%BeginExpansion
$>$%
%EndExpansion
0.5 mA photocurrent or $10^{16}$ photons/s) or by photon counting detectors
and subsequent differencing schemes (hardware or software) at much attenuated
intensity of less than $10^{-12}$ W. The calculation of the differential
signal based on splitting the wavefront of the wave modes predicts near-zero
($\ll\bar{N}$) noise variance. In sharp contrast, the variance $V(\Delta
N)=\bar{N}_{c}+\bar{N}_{d}=\bar{N}$ is predicted by the definite completion of
the quantum measurement and the reduction of the state at the individual
intensity detectors, without the wave modes of quantum vacuum.\emph{ In the
differential detection experiments, both in the BHD scheme and in the BWDD
scheme, there is no role of physical significance for the differencing
operation after the intensity detectors}.%

%TCIMACRO{\FRAME{ftbpFU}{4.611in}{3.1125in}{0pt}{\Qcb{The noise variance
%measured in the balanced wavefront-division measurement, plotted as a function
%of the (balanced) optical power in each detector. Noise variance scales with
%the optical power (dotted line), similar to the noise in the homodyne
%measurement, indicating that the wave mode of quantum vacuum is not the source
%of the quantum noise.}}{\Qlb{f-four}}{fig4.eps}%
%width 4.611in;  height 3.1125in;  depth 0pt;  original-width 4.8003in;
%original-height 3.232in;  cropleft "0";  croptop "1";  cropright "1";
%cropbottom "0";  filename 'fig4.eps';file-properties "XNPEU";}}}%
%BeginExpansion
\begin{figure}
[ptb]
\begin{center}
\includegraphics[
height=3.1125in,
width=4.611in
]%
{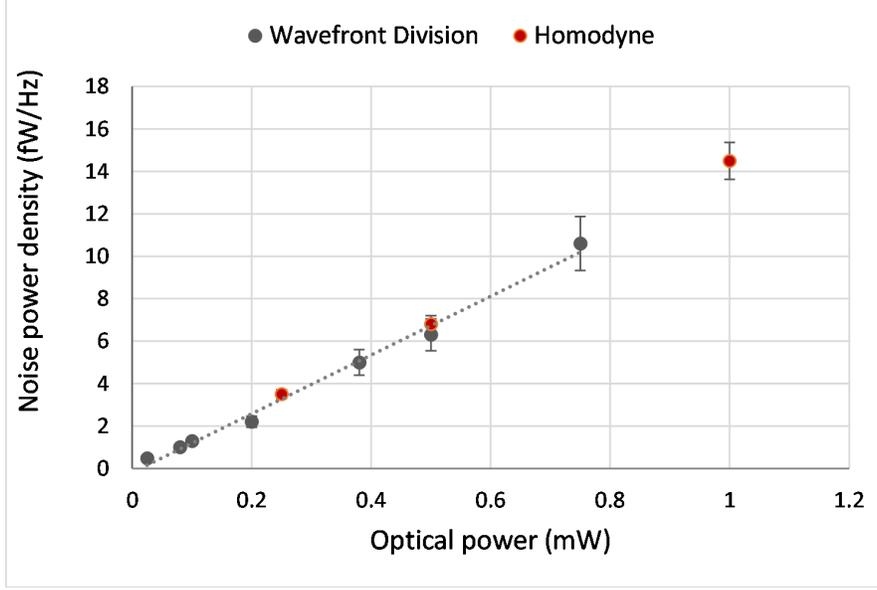}%
\caption{The noise variance measured in the balanced wavefront-division
measurement, plotted as a function of the (balanced) optical power in each
detector. Noise variance scales with the optical power (dotted line), similar
to the noise in the homodyne measurement, indicating that the wave mode of
quantum vacuum is not the source of the quantum noise.}%
\label{f-four}%
\end{center}
\end{figure}
%EndExpansion

The dark current of commercial balanced detectors is in the range $1-5$
pA/$\sqrt{Hz}$, below 10 MHz. For example, a NewFocus-1807 model balanced
detector has a noise equivalent power of 3.3 pW/$\sqrt{Hz}$ in the range DC-10
MHz and the model 2007 (Nirvana) balanced detector has the noise figure of 1.7
pA/$\sqrt{Hz}$, in the reduced bandwidth of 125 kHz. The quantum shot noise
exceeds the detector dark noise only when the optical power is about 0.05 mW (%
%TCIMACRO{\TEXTsymbol{>}}%
%BeginExpansion
$>$%
%EndExpansion
$10^{14}$ photons/s), corresponding to photocurrent fluctuations of a few
pA/$\sqrt{Hz}$. The optical quantum noise measured at 110 kHz in a wavefront
division differential detection experiment using the NewFocus-2007 detector,
in the scheme shown in figure \ref{f-three}A, is plotted in the figure
\ref{f-four} (the noise spectrum is flat within 2 dBm/Hz in the range 100
kHz-130 kHz). The detector shot noise, without any light input, is measured to
be below -126 dBm/Hz (%
%TCIMACRO{\TEXTsymbol{<}}%
%BeginExpansion
$<$%
%EndExpansion
0.25 fW/Hz). With 0.02-0.8 mW of light in each half of the wavefront from a
He-Ne laser, the light beams are balanced to get a differential voltage output
less than 10 mV, corresponding to a differential power below 1 $\mu$W. The
differential scheme cancels the relatively large (1\% rms) laser intensity
noise by a factor of nearly $10^{5}$. The residual noise is monitored in the
100 kHz-150 kHz bandwidth with a spectrum analyzer. The near-zero electrical
(voltage) output of the differential circuit is connected to the spectrum
analyzer and voltmeter, to directly measure the spectral density and rms value
of the residual noise variance of the differential signal, $V(\Delta
N(t))=\left\langle \left(  \Delta N-\overline{\Delta N}\right)  ^{2}%
\right\rangle \propto\left\langle \left(  \Delta i-\overline{\Delta i}\right)
^{2}\right\rangle $.

The noise variance in BWDD scheme scales linearly with the optical power in
each half of the wavefront, in spite of the subtraction of equal portions of
the same wavefront, without the phase factor $\pi$ \ that is characteristic of
the homodyne differential measurement. Data from the balanced homodyne
measurement is also shown for comparison, in the range 0.25-1 mW. Variances in
both cases are similar and scale as the average optical power. The subtraction
scheme does not cancel the quantum noise in the BWDD scheme, directly
contradicting the standard calculation in section IIIC. This indicates that
the quantum noise in each half of the optical wavefront is independent, as
predicted for the quantum state reduction at the detector, ruling out the
physical reality of the fluctuating wave modes of the quantum vacuum.

\begin{figure}
[ptb]
\begin{center}
\includegraphics[
height=1.5774in,
width=2.6368in
]%
{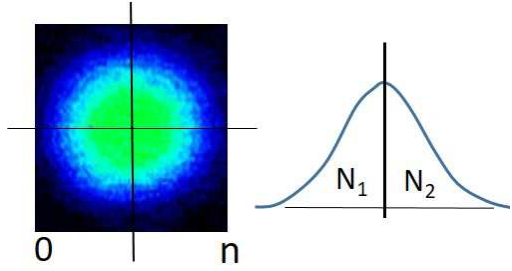}%
\caption{Direct wavefront-division differential measurement of an optical beam
to study quantum noise. }%
\label{f-five}%
\end{center}
\end{figure}

There is another class of measurements that are effectively identical to
balanced wavefront-division differential measurements. To understand this,
consider the measurement of beam parameters of the optical beam from a single
coherent laser beam of any intensity. This can be done by sub-saturation
imaging with a CCD camera. We note at once that the task of determining the
`centre' of the beam, or `centroiding', involves just taking the difference
signal between `two halves' on any chosen diameter and averaging (figure
\ref{f-five}). The sum of pixels on each half is equivalent to one detector.
Since the image of the beam is two-dimensional, the division can be done in
many ways. This difference signal has quantum noise, exactly like the quantum
noise observable in the experiment indicated in fig. \ref{f-three}.%

%TCIMACRO{\FRAME{ftbpFU}{2.6368in}{1.5774in}{0pt}{\Qcb{Direct
%wavefront-division differential measurement of an optical beam to study
%quantum noise. }}{\Qlb{f-five}}{fig5.eps}%
%width 2.6368in;  height 1.5774in;  depth 0pt;  original-width 3.2412in;
%original-height 1.9278in;  cropleft "0";  croptop "1";  cropright "1";
%cropbottom "0";  filename 'fig5.eps';file-properties "XNPEU";}}}%
%BeginExpansion
%EndExpansion

Another example is the determination of centroid of the light through a slit,
as in an optical lever that measures angular changes. Still another is the
localization of the telescope image of stars. While the total width of the
image is determined by diffraction limit of the imaging system, the centroid
error is ultimately defined by the quantum noise.

\emph{Centroiding is the operation of finding out the pixel location that
equally divides the total number of photons in the detector (}$N(t)$\emph{ or
}$i(t)$\emph{) into two bins}. If one repeats this, the pixel location
fluctuates (given sufficient spatial resolution) even if only quantum noise is
present. For our task of determining the quantum noise, the exact location of
the centroid is not critical. We consider the situation in which the balance
of average number of photons is better than $1\%$. Our target is the
variations in $\Delta N(t)$ and adjusting $\Delta N(t)$ to exactly zero is not
required. A cooled EMCCD camera (detector) has high quantum efficiency and
nearly single photon sensitivity. The optical beam can be expanded so that a
large part of the wavefront (%
%TCIMACRO{\TEXTsymbol{>}}%
%BeginExpansion
$>$%
%EndExpansion
$98\%$) illuminates the sensor. The difference in the total counts from pixels
on the two sides of the fiducial pixel is
\begin{equation}
\Delta N(t)=%
%TCIMACRO{\dsum \limits_{0}^{n/2}}%
%BeginExpansion
{\displaystyle\sum\limits_{0}^{n/2}}
%EndExpansion
N_{i}-%
%TCIMACRO{\dsum \limits_{n/2}^{n}}%
%BeginExpansion
{\displaystyle\sum\limits_{n/2}^{n}}
%EndExpansion
N_{i}%
\end{equation}
where each $N_{i}$ is $\alpha^{\dag}\alpha\simeq\alpha\alpha^{\ast}$ that we
calculated earlier. In all experiments dealing with wavefront division, it is
necessary to make sure that the spatial (angular) fluctuation of the whole
beam is small enough to be negligible. Since the whole beam with its quantum
fluctuations is described by the amplitude $\alpha=a+\delta a_{1}+i\delta
a_{2}$, each $N_{i}$ is proportional to $\alpha\alpha^{\ast}$ (with factors
for the pixel area, amplification factor and a Gaussian beam factor etc.)
\ Therefore, $\Delta N(t)\simeq0$ and its variance is also zero, according to
the equations \ref{Del-N} and \ref{Var-N}. \ However, centroiding experiments
that measure the beam profiles of weak laser beams show the fluctuations in
$\Delta N(t)$ with variance
\begin{equation}
V(\Delta N)\simeq N
\end{equation}

The physical problem of locating the centroid of the optical image of an
illuminated slit, as in autocollimating optical levers to measure angular
changes, is similar. The angular shift is measured by tracking the centroid of
the image of the slit, which in turn is the balanced differencing operation on
the light intensities of the two `halves' of the image. So, a BWDD scheme is
operational in such devices. Optical levers operated at their quantum
sensitivity limit \cite{Jones-review} already indicate that the observed noise
floor is in conflict with the hypothesis that the quantum vacuum mode is the
source of quantum noise. Take the case of the instrument with an
experimentally determined sensitivity limit of $<10^{-10}$ radians/$\sqrt
{\text{Hz}}$, with total light detected at the photodiode of about $1$ $\mu
W$, or about $10^{13}$ photons/s \cite{Jones-review}. \ The diffraction
limited image of the slit had an angular width of $2\times10^{-4}$ radians.
With $\sqrt{V(\Delta N)}=\sqrt{N}\simeq3\times10^{6}$, the quantum limited
sensitivity of such an optical lever is then $\delta\theta>2\times
10^{-4}/3\times10^{6}\simeq7\times10^{-11}$ radians/$\sqrt{\text{Hz}}$. In
contrast, the prediction for the quantum noise in the difference signal,
assuming that its source is the fluctuating quantum vacuum wave mode, is
$V(\Delta N)\simeq0$ . Since other noise sources are controlled better in the
device, the experimental results of shot-noise limited sensitivity of the
optical lever reject the hypothesis of real wave modes of the quantum vacuum.
Instead, they support the picture of local quantum state reduction of the
superposition of number states at each pixel, at the instant of square-law
(intensity) detection.

It is instructive to compare results on the quantum noise in optics with the
quantum noise in the direct detection of \ atoms in a coherent beam of atomic
Bose-Einstein condensate (BEC) of metastable Helium atoms (which allows such
direct detection and counting of the atoms on a delay line detector due to the
nature of spontaneous ionization of the He* atoms on contact) \cite{Aspect-He}%
. There is quantum zero-point energy in the dynamics of the atomic BEC, finite
and proportional to the number of atoms. Atomic BEC is in a coherent state,
but it has no associated real wave modes of quantum vacuum in space. All
zero-point energy of the atomic BEC is associated with the atoms and it is
truly zero when there are no atom in the beam. However, the quantum noise in
the difference in the number of atoms on either side of a `centroid' has the
variance, $V(\Delta N)\simeq N$. Clearly, this quantum noise is not related to
any wave modes of quantum vacuum.

\section{Balanced Homodyne Scheme and Michelson Interferometry}

\begin{figure}
[ptb]
\begin{center}
\includegraphics[
height=1.9817in,
width=3.0959in
]%
{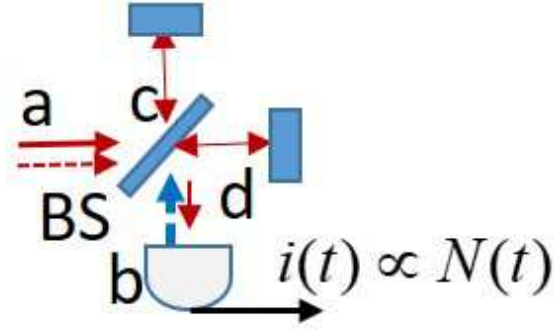}%
\caption{The quantum measurement in the Michelson interferometer. }%
\label{f-six}%
\end{center}
\end{figure}

The common passive optical element of significance in a balanced homodyne
measurement and Michelson interferometry is the symmetric (50:50) BS. The
detection is entirely different though. In homodyne, there are two detectors
that \emph{complete} the quantum measurements and a differencing circuit (hard
or soft) produces the final difference signal. In Michelson interferometry,
the input light  is divided into two paths by the BS and two end mirrors
reflect the beams back to the BS, preserving the average phase after the
double-pass through BS. This is equivalent to a \emph{differencing operation
on the beam amplitudes.} This is then detected with a single square law
detector that completes the quantum measurement (figure \ref{f-six}). Compared
to the homodyne detection, there are significant differences in details of the
quantum noise that are important in the application of squeezed light in
sub-shot noise metrology.%

%TCIMACRO{\FRAME{ftbpFU}{3.0959in}{1.9817in}{0pt}{\Qcb{The quantum measurement
%in the Michelson interferometer. }}{\Qlb{f-six}}{fig6.eps}%
%width 3.0959in;  height 1.9817in;  depth 0pt;  original-width 1.9228in;
%original-height 1.2204in;  cropleft "0";  croptop "1";  cropright "1";
%cropbottom "0";  filename 'fig6.eps';file-properties "XNPEU";}}}%
%BeginExpansion
%EndExpansion

The difference signal in the homodyne scheme with a single beam at port `c' is
the interference term $\Delta N(t)=c^{\dag}c-d^{\dag}d=ab^{\dag}+a^{\dag
}b\simeq2a\delta b_{1}$ with variance $4a^{2}\delta b_{1}^{2}$. The radiation
pressure noise in a Michelson interferometer has the same feature since the
light pressure on each mirror is proportional to the number of photons
reflecting off the mirror and the interferometer optical noise comes from the
fluctuating differential motion of the two mirrors; the common mode
displacement does not lead to an optical intensity modulation at the detector.
However, this view is simplistic. Consider a Michelson interferometer with one
of the mirrors anchored, with its free motion arrested. The interferometer is
still sensitive to the gravitational wave strain, albeit with a reduction in
sensitivity by a factor of two. However, now the radiation pressure noise is
related to $\delta N(t)=c^{\dag}c-\left\langle c^{\dag}c\right\rangle =a\delta
a_{1}+a\delta b_{1}$ with variance
\begin{equation}
V(\delta N)=a^{2}\left\langle \left(  \delta a_{1}^{2}+\delta b_{1}%
^{2}+2\delta a_{1}\delta b_{1}\right)  \right\rangle \simeq a^{2}\left\langle
\delta a_{1}^{2}+\delta b_{1}^{2}\right\rangle =\frac{1}{2}V(\Delta N)
\end{equation}
It is not natural or logical to claim that just by holding one of the mirrors
the radiation pressure noise became equal contribution of vacuum noise in both
modes $a$ and $b$, whereas with mirrors left free, the entire quantum noise
was solely from the vacuum noise of $b$ entering the open port. In both cases,
there is no constraint on the operation point, which can continue to be the
dark fringe. This becomes important in the context of squeezed light injection
in the open port for reducing radiation pressure noise at low frequencies.
With both mirrors free, the wave-mode picture allows the possibility of
arbitrary reduction of radiation pressure noise by injecting light squeezed by
the factor $\exp(-s)$,
\begin{equation}
V(\Delta N)\simeq4e^{-2s}a^{2}\delta b^{2}%
\end{equation}
whereas the theory allows only limited noise reduction when one of the mirrors
in held rigid,
\begin{equation}
V(\delta N)\simeq a^{2}\delta a^{2}+e^{-2s}a^{2}\delta b^{2}%
\end{equation}

Currently, the advances in quantum noise reduction in gravitational wave
detectors based on Michelson interferometers are based on the notion that the
quantum noise in the Michelson signal is similar to the quantum noise in the
homodyne signal, which is the cross interference term between the coherent
amplitude (or LO) and the quantum noise in the vacuum mode entering the open
port, $2a\delta b$. The quantum noise is in two quadratures, $\delta b_{1}$
`in phase' with the LO amplitude and $\delta b_{2}$ that is orthogonal. While
the cross term in the homodyne signal is $a\delta b_{1}$, it is $a\delta
b_{2}$ in the Michelson interferometer signal, operated near the dark fringe.
This is because the intensity fluctuations near the dark fringe (no light
output at port `b') are due to the fluctuating \emph{phase difference}, rather
than amplitude fluctuations, of the light in the two arms. Therefore, one
argues that replacing the rogue quantum vacuum mode with the squeezed vacuum,
with smaller $\delta b_{2}$, at the cost of increased $\delta b_{1}$, reduces
the noise in the relevant quadrature \cite{CavesPR81}. This expectation turns
out to be correct, but the detailed physical reason does not involve the wave
modes of the quantum vacuum and their fluctuations; the comparison of the BWDD
scheme with the balanced homodyne scheme already rules out the reality of the
wave modes of quantum vacuum. \ The correct interpretation, in which the
quantum state reduction of the superposition of the photon number states
happens at the intensity detector where the quantum measurement is completed,
will be discussed in a subsequent paper. While the use of the wave modes is
very convenient in calculations, just like in the calculation of the Casimir
effect etc., the physical reality of such modes is in conflict with both
laboratory experiments and cosmology. Hence a consistent and complete
description in terms of photon statistics during quantum state reduction at
measurement, without referring to the fluctuating wave modes, is necessary and possible.

\section{Summary}

I have examined two kinds of balanced differential optical measurement
involving the difference of intensities recorded by two photodetectors. The
first is the balanced amplitude-division homodyne detection (BHD) that uses a
symmetric BS and the second is the balanced wavefront-division detection
(BWDD) in which the BS is absent. The two beams in each scheme are detected
with separate detectors and the difference signal and its quantum fluctuations
are compared to determine the physical reality and the contribution of the
hypothetical quantum vacuum mode to the factually observed quantum noise. The
relevant theoretical expressions were derived from first principles, for each
experiment. It was shown that the predictions for the variance of the
difference signals are very different in the BHD scheme and the BWDD scheme.
In BHD, the interference from fluctuations in the quantum vacuum mode from one
input port of the BS entirely cancels in the subtraction, while the
fluctuations of the vacuum mode through the second port survive as the cross
interference term with the carrier and manifest as the quantum noise of the
homodyne signal. This conventional picture and the associated theory seem
consistent with the results of the homodyne experiments.

However, the \ BWDD scheme provides a decisive counter-point because the
interference of the fluctuations disappears totally in the subtraction without
a beam splitter and there is no cross interference term. Therefore, if the
measured quantum noise in the BWDD is very different from the prediction of
vanishing residual noise, then the experiment decisively rules out the reality
of the quantum vacuum mode and its role in the observed quantum noise. An
experiment in which a balanced pair of photodetectors were used to measure the
differential intensity and its variance from splitting the wavefront showed
that the quantum noise variance scales linearly with optical power, instead of
cancelling in the subtraction. Several experimental schemes that are
equivalent to the BWDD scheme are examined to confirm this result. This
decisively rejects the physical reality of the wave modes of quantum vacuum.
Instead, the reduction of the superposition of particle number states to a
specific but random number state in each quantum measurement is the sole
consistent description and the preferred interpretational choice of the
quantum noise. This is also consistent with observational cosmology and its
general relativistic description, since there is no divergent ZPE density in
this picture. I have derived the quantum noise in this alternate picture in
all cases of optical detection. It correctly predicts equal quantum noise in
both detection schemes, BHD and BWDD.

These results are corroborated by the recent reformulation of general dynamics
and quantum mechanics in terms of a universal wave equation for the `waves of
action', rather than for the hypothetical matter-energy waves
\cite{Unni-RQM-arxiv}. This modification of dynamics that promotes Hamilton's
equation to an action-wave equation reproduces quantum interference,
correlations and uncertainty noise, without the divergent ZPE because the
action waves carry `action' and not energy or momentum. The uncertainty
principle is recast as $\Delta S\geq\hbar$. Schr\"{o}dinger equation and its
wavefunction pertain to statistical averages over an ensemble of dynamical
histories and not single quantum history, contrary to the prevailing
understanding of quantum mechanics. The physical basis of the use of squeezed
light in quantum metrology in the light of these results remains to be
discussed, especially in the context of the Michelson interferometer. More
results and discussion on squeezed light quantum metrology without quantum
vacuum modes will be presented in a subsequent paper.

\bigskip
\noindent{\bf Acknowledgements}

I have benefitted from conversations on the wave modes of quantum vacuum and
the Casimir effect with G. Rajalakshmi, D. Suresh, R. Cowsik, N. D. Haridass,
K. A. Milton, S. Reynaud, and N. Jetty. I thank N. Jetty and S.
Sankaranarayanan for several discussions on optical quantum noise and its
calculation and interpretation in quantum optics. P. G. \ Rodrigues and P. V.
Sudersanan helped in setting up the experiments.

\end{document}